\documentclass{elsart3p}

\usepackage{natbib}
\input epsf
\usepackage{epsfig}
\usepackage{amssymb}
\usepackage{amsmath}
\def\plotone#1{\centering \leavevmode
\epsfxsize=2 \columnwidth \epsfbox{#1}}
\def\plottwo#1#2{\centering \leavevmode
\epsfxsize=\columnwidth \epsfbox{#1} \hfil
\epsfxsize=\columnwidth \epsfbox{#2}}

\newcommand\lsim{\mathrel{\rlap{\lower4pt\hbox{\hskip1pt$\sim$}}
        \raise1pt\hbox{$<$}}}
\newcommand\gsim{\mathrel{\rlap{\lower4pt\hbox{\hskip1pt$\sim$}}
        \raise1pt\hbox{$>$}}}

\begin{document}

\begin{frontmatter}

\title{Cosmological Physics with Black Holes\\ (and Possibly White Dwarfs)}


\author{Kristen Menou, Zoltan Haiman}

\address{Department of Astronomy, Columbia University, 550 West 120th Street, New York, NY 10027, USA}

\author{Bence Kocsis}

\address{Harvard-Smithsonian Center for Astrophysics, 60 Garden Street, Cambridge, MA 02138, USA}

\thanks{K. Menou gratefully acknowledges more than a decade of
exchanges with Jean-Pierre Lasota on the physics of black holes (and
definitely white dwarfs).}

\hspace{10 mm}

\begin{abstract}
% Text of abstract
The notion that microparsec-scale black holes can be used to probe
gigaparsec-scale physics may seem counterintuitive, at first. Yet, the
gravitational observatory {\it LISA} will detect
cosmologically-distant coalescing pairs of massive black holes,
accurately measure their luminosity distance and help identify an
electromagnetic counterpart or a host galaxy.  A wide variety of new
black hole studies and a gravitational version of Hubble's diagram
become possible if host galaxies are successfully identified.
Furthermore, if dark energy is a manifestation of large-scale modified
gravity, deviations from general relativistic expectations could
become apparent in a gravitational signal propagated over cosmological
scales, especially when compared to the electromagnetic signal from a
same source. Finally, since inspirals of white dwarfs into massive
black holes at cosmological distances may permit pre-merger
localizations, we suggest that careful monitoring of these events and
any associated electromagnetic counterpart could lead to
high-precision cosmological measurements with {\it LISA}.
\end{abstract}


\end{frontmatter}

% main text
\section{Introduction}
\label{sec:intro}

Essentially all astronomical measurements are performed via
electromagnetic waves. The availability of accurate gravitational wave
measurements within the next decade or so will thus be a significant
development for astronomy. In particular, since the propagation of
photons and gravitons could differ at a fundamental level,
gravitational waves emitted by cosmologically-distant ``space-time
sirens,'' such as coalescing pairs of massive black holes, could be
used as valuable new probes of physics on cosmological scales.

Black holes with masses $\gsim 10^6 M_\odot$ are present at the center
of numerous nearby galaxies \citep[e.g.][]{kr95,mag98}. As such
galaxies collide over cosmic times, their central black holes
coalesce, releasing $\gsim 10^{58}$~ergs of binding energy in the form
of gravitational waves (hereafter GWs). To measure the GWs emitted by
these cosmologically-distant space-time sirens, ESA and NASA will
build the Laser Interferometer Space Antenna, LISA\footnote{{\tt
http://lisa.nasa.gov/}}.

\begin{figure}
\centering \leavevmode
\epsfxsize=\columnwidth \epsfbox{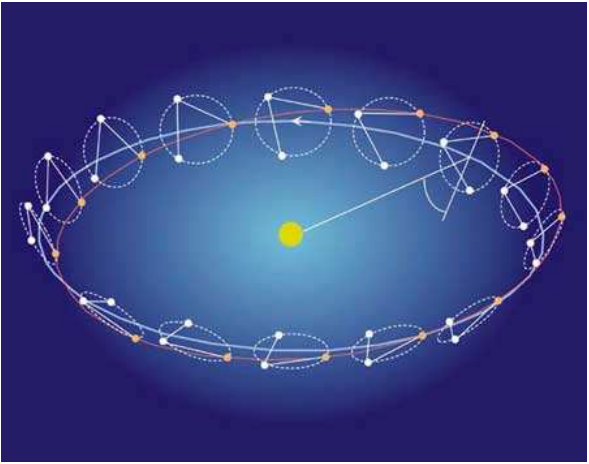}
\caption{Orbit of the LISA 3-base space interferometer, trailing the
Earth around the Sun. A typical coalescing black hole pair observed
for a year is well-modulated by the interferometer and thus accurately
localized on the sky. [Credit: {\tt http://lisa.nasa.gov/}]}
\label{fig:one}
\end{figure}

GWs emitted by black hole binaries have the unfamiliar property of
providing a direct measure of the luminosity distance, $D_L$, to the
black holes, without extrinsic calibration. Owing to the highly
coherent nature of GW emission \citep{schu86}, the amplitude (or
strain), $h_{+\times}$, frequency, $f$, and frequency derivative,
$\dot f$, of the leading order (quadrupolar) GW inspiral signal scale
as
\begin{eqnarray} 
h_{+\times} (t) & \propto & \frac{\left[ (1+z) M_c \right]^{5/3}
f^{2/3}}{D_L}, \\ \dot f (t)& \propto & \left[ (1+z) M_c \right]^{5/3}
f^{11/3},
\end{eqnarray}
 where $+\times$ represents the two transverse GW polarizations, $M_c=
(m_1 m_2)^{3/5} / (m_1+m_2)^{1/5}$ is the black hole pair ``chirp''
mass and $z$ its redshift. Provided the GW source can be reasonably
well localized on the sky, an extended observation of the chirping
signal leads to precise measurements of $h_{+\times}$, $f$, $\dot f$
and thus $D_L$, independently. As illustrated in Fig.~\ref{fig:one},
LISA's orbital configuration allows for a ``triangulation'' of GW
sources on the sky, to within a solid angle $\delta \Omega \sim
1$~deg$^2$ typically \citep{cut98,vec04}. This permits very accurate
measurements, e.g. distances with errors $\delta D_L / D_L < 1\%$ at $z \lsim
2$ typically \citep{cut98,hug02,vec04,lh06}. Masses are independently
determined to very high accuracy (typically $\ll 1\%$; e.g.,
\citealt{hug02})

\section{Post- and Pre-Merger Localizations}

In principle, the same sky localization that helps determine the
distance to a source accurately can be used to find the host galaxy of
a pair of merging black holes seen by LISA. The secure identification
of the host galaxy would enable a wide variety of new galactic black
hole studies (see \S\ref{sec:newBH}).

Initially, the prospects for finding the host galaxy of a pair of
merging black holes were considered to be poor, simply because of the
large number of galactic candidates located in the $\delta \Omega \sim
1$~deg$^2$ LISA sky error-box \citep[e.g.,][]{cut98,vec04} Recently,
however, this possibility has been reconsidered, with more optimistic
conclusions \citep{hohu05,koc06,koc07}.

\begin{figure}
\centering \leavevmode
\epsfxsize=\columnwidth \epsfbox{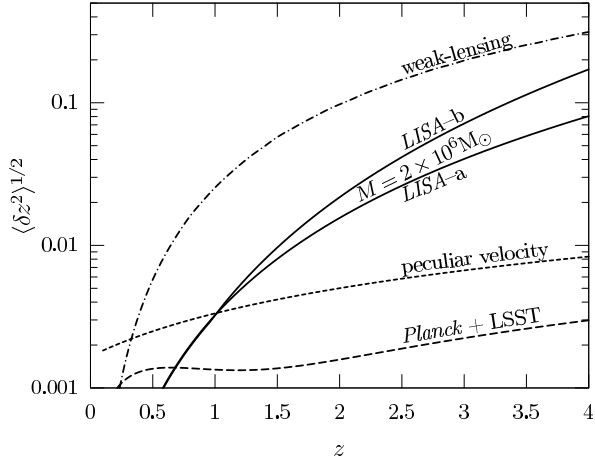}
\caption{Contributions to the error on the inferred redshift of an
 electromagnetic counterpart to a {\it LISA} coalescence event, as a
 function of redshift $z$, for $m_1=m_2=10^6 M_\odot$. The intrinsic
 {\it LISA} error on the luminosity distance, $d_L$, is shown for two
 representative cases (a \& b, solid lines). Errors due to the
 peculiar velocity of the source (for $v=500~{\rm km~s^{-1}}$;
 short--dashed line), uncertainties on the background cosmology
 (long--dashed line), and errors due to weak lensing magnification
 (dash--dotted line) are also shown (see \citealt{koc06} for
 details).}
\label{fig:two}
\end{figure}

Given a cosmology, it is possible to translate the accurate luminosity
distance measurement to the GW source into a narrow redshift slice in
which the host galaxy must be located \citep{hohu05,koc06}. Various
contributions to the redshift errors that arise in performing this
conversion are shown in Fig.~\ref{fig:two}, for a representative
equal-mass binary, as a function of the GW source redshift
\citep{koc06}.  At redshifts $z \gsim 0.25$, where most black hole
binary sources are expected to be found, weak lensing errors due to
line-of-sight inhomogeneities (on top of the smooth average cosmology)
are the main limitation to an accurate determination of the redshift
slice in which the host galaxy ought to be located.

\citet{koc06} have studied in detail the possibility that the
three-dimensional information available (sky localization + redshift
slice) could be used to single out a quasar, or any other unusually
rare object (such as a star-bust galaxy), in the LISA error box, after
coalescence. Finding such a statistically rare object post-merger would
make it a good host galaxy candidate for the newly-coalesced pair of
black holes.

However, it maybe much more advantageous to use a pre-merger strategy
to identify the host galaxy of a pair of coalescing black holes seen
by LISA. Indeed, one can use near real-time GW information on the sky
localization, in combination with the accurate timing of the inspiral
event, to predetermine well in advance where on the sky the merger is
located. A unique host galaxy identification could then proceed
through coordinated observations with traditional telescopes, by
monitoring in real time the sky area for unusual electromagnetic
emission, as the coalescence proceeds.

A variety of mechanisms exist through which disturbed gas in the
vicinity of black hole pairs will power electromagnetic emission
during and after coalescence
\citep{armnat02,milphi05,dot06,boph07,macfadyen07}.  For example, at
the time of coalescence, $\gsim 10^{53}$~ergs of kinetic energy are
delivered to the recoiling black hole remnant and its environment, for
typical recoil velocities $\gsim 100$ km/s \citep[e.g.,][]{baker06,baker07,camp07,herr07,sb07}. This may lead to detectable signatures
\citep{lippai08,milphi05} and permit the coincident identification of
a unique host galaxy. The detailed nature of such electromagnetic
counterparts remains largely unknown, however.

\begin{figure}
\plottwo{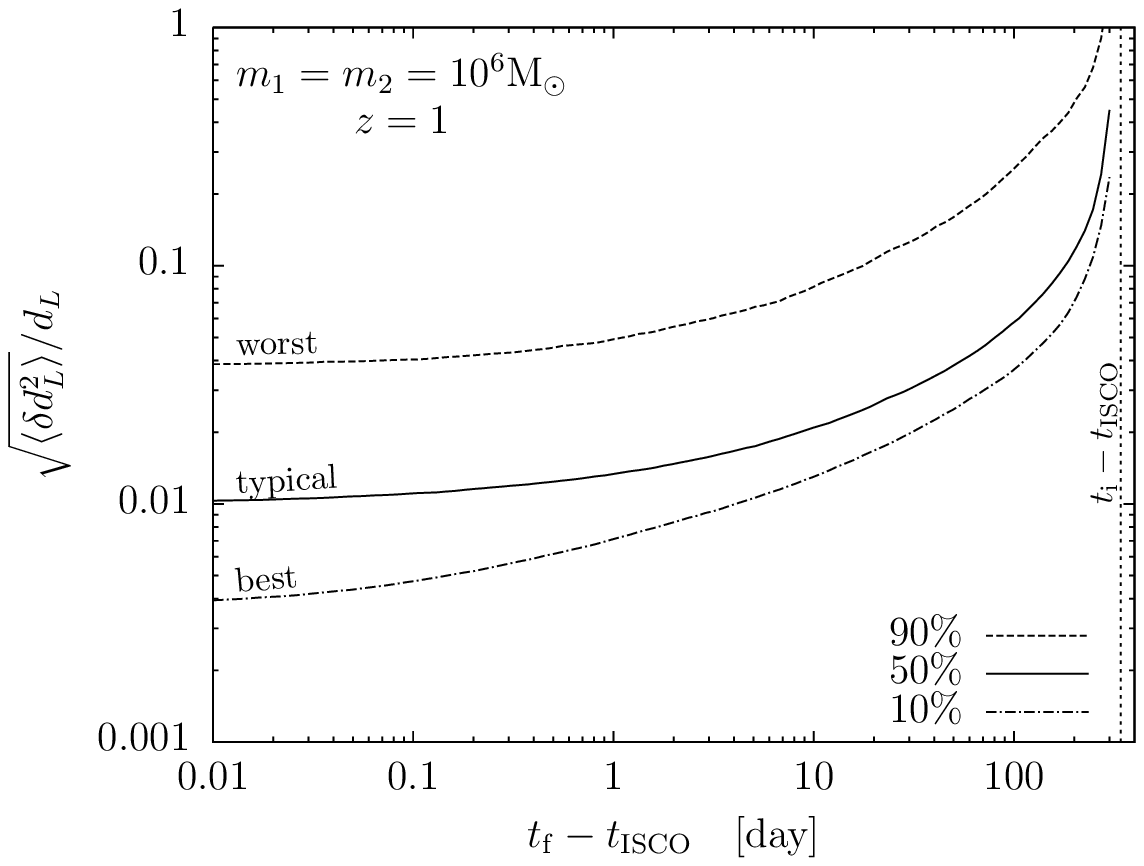}{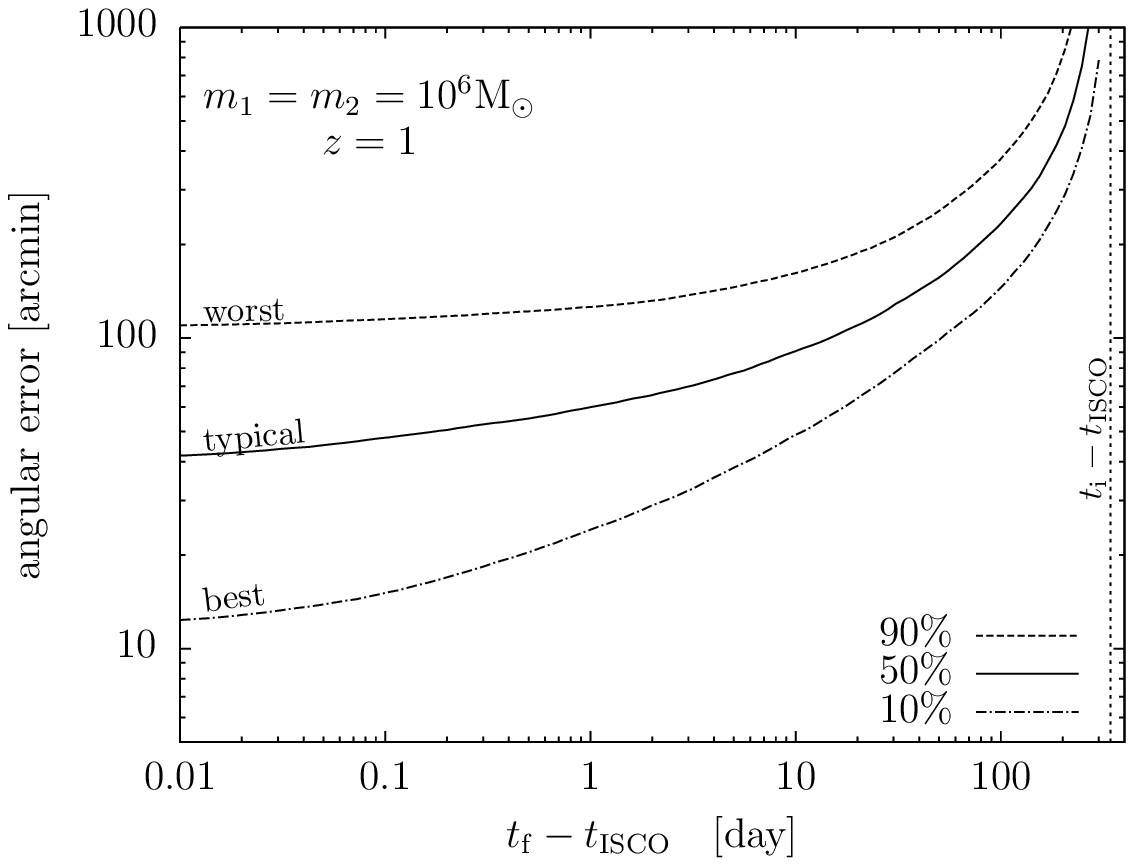}
\caption{Evolution with pre-merger look--back time,
$t_{\mathrm{f}}$-$t_{\rm ISCO}$, of LISA source localization errors,
for $M=2\times 10^6 M_\odot$ and $z=1$. The top panel shows luminosity
distance errors and the bottom panel shows sky position angular errors
(equivalent diameter, $2\sqrt{ab}$, of the error ellipsoid). Best,
typical, and worst cases for random orientation events represent the
$10\%$, $50\%$, and $90\%$ levels of cumulative error distributions,
respectively.  Errors for worst case events effectively stop improving
at a finite time before merger, even though the signal-to-noise ratio
accumulates quickly at late times. Errors for best case events
(especially the minor axis) follow the signal-to-noise ratio until the
final few hours before merger (see \citealt{koc07} for details)}
\label{fig:three}
\end{figure}

To a large extent, LISA's ability to localize a long-lived source on
the sky is related to the GW signal being modulated as a result of the
detector's revolution and change of orientation when the constellation
orbits around the Sun (Fig.~\ref{fig:one}). Even though most of the GW
SNR accumulates during the final stages of inspiral/coalescence for
typical GW sources, reasonably good information on sky localizations
must be available well before final coalescence since this information
accumulates slowly, over the long signal modulation (orbital)
timescale. Because of significant cross-correlations between sky
localization and distance errors, it turns out that this argument is
also largely valid for luminosity distance errors
\citep{hohu05,koc06,koc07}.

Figure~\ref{fig:three} shows the pre-merger time evolution of
luminosity distance and angular sky localization errors for a
representative black hole pair at $z=1$. Errors improve quickly at
early times but their evolution slows down considerably at late times.
According to both panels, even accounting for random orientations of
various source and detector angles (shown as best, typical and worst
cases), significant information is available days to weeks prior to
the final coalescence \citep{koc07}. Including black hole spins in the
analysis has been shown to result in significant improvements on the
errors during the last few days to hours prior to coalescence
\citep{lh08}.

\begin{figure}
\centering \leavevmode
\epsfxsize=\columnwidth \epsfbox{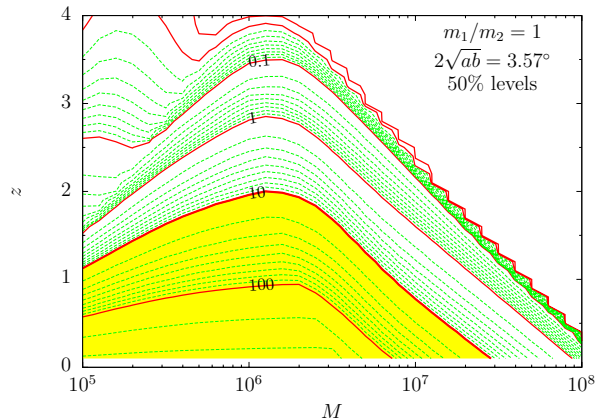}
\caption{Contours of advance warning times in the total mass ($M$) and
redshift ($z$) plane for equal-mass black hole binaries ($m_1/m_2=1$).
The contours trace the look--back times at which the equivalent radius
($2\sqrt{ab}$) of the LISA localization error ellipsoid first reaches
an LSST-equivalent field-of-view ($3.57^{\circ}$). These contours
correspond to the $50\%$ level of cumulative distributions for random
orientation events.  The contours are logarithmically spaced in days
and $10$ days is highlighted with a thick line (see \citealt{koc08}
for details).}
\label{fig:four}
\end{figure}

With the expected availability, by the time LISA is operational, of
sensitive large field-of-view (FOV) astronomical instruments for weak
lensing and supernova studies, it becomes interesting to estimate the
amount of time prior to merger during which the LISA sky localization
falls within the FOV of such an instrument. When this happens,
continuous monitoring of the designated sky area, until final
coalescence, becomes possible. \citet{koc07} have performed a detailed
analysis of this possibility, using LSST and its $10$~deg$^2$ FOV as a
reference. Figure~\ref{fig:four} shows results for representative
equal-mass binaries, as a function of their total mass and
redshift. The various contours show that prospects for electromagnetic
monitoring days to weeks before the coalescence are good for sources
at redshifts $z \lsim 2$--$3$. Monitoring for the best GW sources out
to $z \sim 5$-$7$ may even be possible \citep{koc07}.

\section{New Science with Electromagnetic Counterparts}

\subsection{Galactic Black Hole Astrophysics} \label{sec:newBH}

A large variety of new galactic black hole astrophysics would be
enabled by successful identifications of the host galaxies of
coalescing black hole pairs. We mention only a few possibilities here
and refer the interested reader to \citet{koc06} and \citet{koc08} for
additional discussions.

From the black hole masses, spins and binary orientation, all
accurately constrained by the GW signal, one would be able to study
the physics of the post-merger accretion flow onto the remnant black
hole \citep{milphi05,dot06} with unprecedented accuracy. This would
include precise constraints on the Eddington ratio of the accreting
source, its emission and absorption geometries and possibly its jet
phenomenology.  Similarly, studies of the galactic host might tell us
about the nature (dry/wet) and the timing of the galactic merger that
resulted in the black hole binary coalescence. Finally, measuring
velocity dispersions, $\sigma$, for several host galaxies,
together with the black hole masses known from the GW signal, would
allow us to accurately map the evolution of the $M_{\rm bh}$--$\sigma$
relation with cosmic time, at least for such transitional objects as
the hosts of coalescing black hole pairs.

\subsection{Gravitational Hubble Diagram}

Another consequence of successfully identifying the host galaxies of
coalescing black hole pairs is the possibility to draw a gravitational
Hubble diagram, i.e. one that relates the gravitational luminosity
distances, $D_L$, of these GW sources to the electromagnetic redshifts,
$z$, of their host galaxies.

One of the main interests of a gravitational Hubble diagram arises
from its immunity to common systematics affecting electromagnetic
measurements.  Indeed, a gravitational Hubble diagram, which is based
on gravitational distance measurements with self-calibrated sources,
is not susceptible to any significant bias from absorption, scattering
or reddening of GWs.

In practice, however, the value of such a diagram is limited by
line-of-sight matter inhomogeneities, which generate weak lensing
uncertainties on the gravitational $D_L$ measurement \citep[see
Fig.~\ref{fig:two};][]{hohu05,koc06,dal06,gunnarsson06,linder08}.
While the lensing effect can in principle be averaged out over many
random lines-of-sights, it may not be possible to do so for coalescing
pairs of massive black holes if LISA merger event rates are modest
\citep[e.g., a few tens per year at $z \lsim
5$;][]{men01,wl03,ses04,mic07}. Weak lensing errors on individual
measurements amount to distance uncertainties ranging from $\delta D_L
/ D_L \simeq 1\%$ at $z=0.5$ to $\delta D_L / D_L \simeq 10\%$ at
$z=5$ \citep[e.g.][]{koc06}, which makes a gravitational Hubble
diagram imprecise even at moderate redshifts. The extent to which LISA
events can be used to draw a meaningful Hubble diagram will thus
depend strongly on the actual distribution of massive black hole
merger events with redshifts and the corresponding efficiency of host
galaxy identifications. As we describe in \S\ref{sec:wd}, however,
white dwarf spiraling into massive black holes may offer a practical
avenue for precision cosmology with LISA.

\subsection{Diagnostics of Modified Gravity}

 The possibility that the accelerated expansion of the Universe
results from a failure of general relativity has fueled much
theoretical work on large scale modifications of gravity over the past
few years.  Since building a satisfactory theory of modified
relativistic gravity is a formidable task, any insight that can be
gained from direct observational constraints on the linearized GW
regime cannot be overlooked. LISA, with its ability to measure the GW
signal from cosmologically-distant sources, may thus be one of our
best probes of modified gravity on cosmological scales \citep{dm07}.

One may expect gravity modifications to contain a new length scale,
$R_c$, beyond which gravity deviates from general relativity. In order
to explain the observed accelerated expansion of the Universe, this
scale is expected to be of the order of the current Hubble radius,
$H_0^{-1}$.  An existence proof of modifications of this type is given
by DGP gravity \citep{DGP,lue06}, a braneworld model with an infinite
extra dimension.

\citet{dm07} discuss the possibility that extra-dimensional leakage of
gravity in DGP-like scenarios may lead to cosmologically-distant GW
sources appearing dimmer than they truly are, from the loss of GW
energy flux to the extra-dimensional bulk.  Indeed, in the presence of
large distance leakage, flux conservation over a source-centered
hypersphere requires that the GW amplitude scales with distance $D$
from the source as
\begin{equation} \label{scaling}
h_{+\times} \propto D^{-(dim-2)/2},
\end{equation}
where $dim$ is the total number of space-time dimensions accessible to
gravity modes. Thus, for $dim \geq 5$, it deviates from the usual $
h_{+\times}(D) \propto 1 /D$ scaling. In principle, black hole merger
events and associated host galaxies could thus reveal the leakage of
gravity over scales of order a few Hubble distances, by comparison to
purely electromagnetic Hubble diagrams, which are immune to such
leakage effects.

\begin{figure*}
\plotone{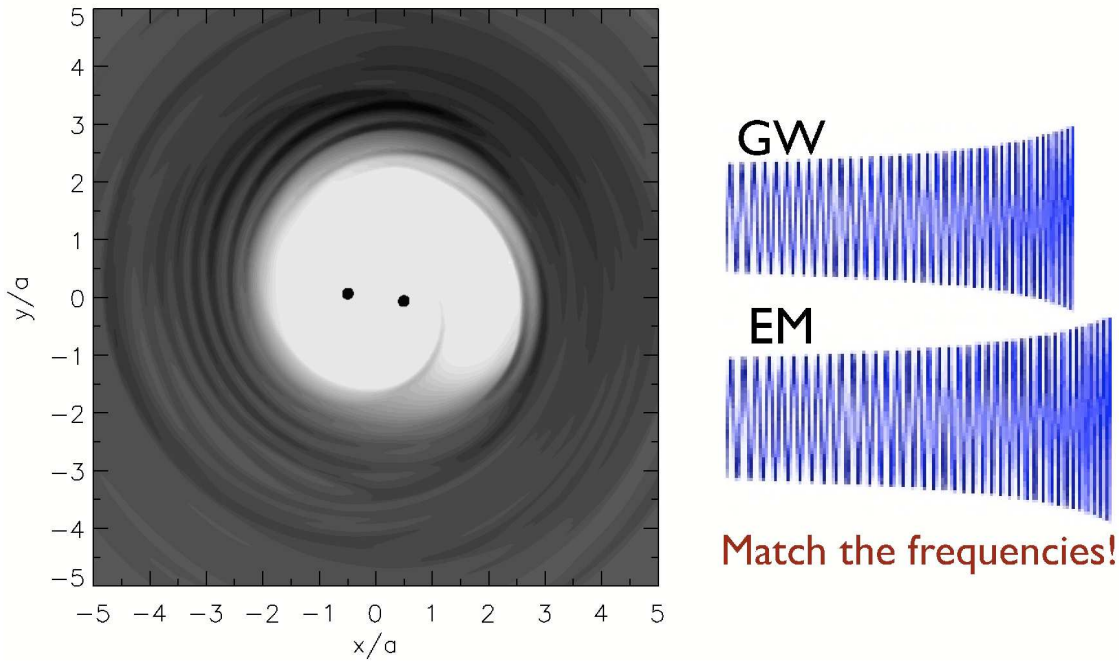}
\caption{[Left] From the numerical simulations of \citet{macfadyen07}:
snapshot of a gaseous disk gravitationally-perturbed by the
time-dependent quadrupolar potential of a central binary. [Right] For
such a system, in the inspiral phase, one may expect an
electromagnetic (EM) source varying in brightness at a frequency
approaching that of the GW signal related to the quadrupolar
perturbation. By matching the frequencies of the EM and GW signals,
one could remove the delay in EM emission at the source, measure
reliable offsets in arrival times between the two signals and possibly
reveal drifts in frequency of the GW signal when it is propagated over
cosmological distances.}
\label{fig:five}
\end{figure*}

This is only one of several possible modified gravity signatures in
the GWs from cosmologically-distant sources \citep{dm07}. Another
class of signatures is related to the GW polarization signal, with
possibly additional polarizations beyond the two transverse
quadrupolar ($+\times$) modes of general relativity
\citep[e.g.,][]{will06}. Signatures also exist in relation to the GW
propagation velocity which, in modified gravity scenarios, can differ
from the speed of light. In this respect, the possibility to time a
cosmological GW, relative to an electromagnetic signal causally
associated with the black hole merger, may offer unique diagnostics of
large-scale modified gravity. This could reveal, for instance, that
the phase of the GW signal deviates from general relativistic
expectations, once propagated over cosmological distances.

\citet{koc08} have explored further the possibilities of measuring
photon and graviton arrival times from a same cosmological source.  A
general difficulty with this approach is that there will be a
systematic and a priori unknown delay in the emission of photons,
relative to the emission of gravitons, since the former must causally
lag behind the perturbing gravitational event. This difficulty could
be overcome if it were possible to calibrate the relative timing of
the photon and graviton signals at the source.

Prior to coalescence, gas present in the near environment of the black
hole binary would be gravitationally perturbed in such a way that it
could radiate a variable electromagnetic signal with a period closely
matching that of the leading-order quadrupolar perturbation induced by
the coalescing binary (see Fig.~\ref{fig:five}). This would help
identify the electromagnetic counterparts of specific GW events.  In
addition, it may be possible to match the variability frequencies of
the electromagnetic and GW signals. The offset in phase between the
Fourier components of the two signals with similar frequencies could
be used to effectively calibrate the intrinsic delay in
electromagnetic emission at the source. Late inspiral and coalescence
can be tracked via the GW signal, so that the relative timing of the
gravitational and electromagnetic signals may be known to within a
fraction of the binary's orbital time.  Any drift in arrival-time with
frequency between the gravitational and electromagnetic chirping
signals, as the source spans about a decade in GW frequency during the
last 2 weeks before merger, could then be attributed to a fundamental
difference in the way photons and gravitons propagate over
cosmological distances.  For instance, such a drift could occur if the
graviton is massive, resulting in a frequency-dependent
propagation velocity \citep[e.g.][]{bbw05,will06,koc08}. This tracking
possibility is illustrated graphically in Fig.~\ref{fig:five}.

Interestingly, while Lorentz invariance has been extensively tested
for standard model fields, Lorentz symmetry could be violated in the
gravity sector, especially on cosmological scales
\citep[e.g.,][]{ceg01,ckr02}. With a good enough understanding of the
source, electromagnetic counterparts to black hole binary mergers may
offer unique tests of Lorentz violations in the gravity sector, via
the opportunity to match and track the gravitational and
electromagnetic signals in frequency and phase. It may be possible, as
the black hole binary decays toward final coalescence, spanning a
range of frequencies, to measure the delays in graviton vs. photon
arrival times as a function of increasing frequency of the chirping
signal. The consistency expected if Lorentz symmetry is satisfied in
the gravity sector could be tested explicitly for gravitons propagated
over cosmological scales.  To have any chance to perform such new
tests of gravitational physics, one will need to identify the
electromagnetic counterparts of coalescing pairs of massive black hole
binaries as early as possible. This may be one of the strongest
motivations behind ambitious efforts to localize these rare, transient
events well before final coalescence.

\section{New Possibilities with White Dwarf Inspirals} \label{sec:wd}

As seen before, a large number of sources must be accumulated to turn
a gravitational Hubble diagram into a high precision tool for
cosmology. White dwarf frequently spiraling into (moderately) massive
black holes may offer unique opportunities in this respect.

An additional goal of the LISA mission is the detection of GWs emitted
by compact objects being captured by massive black holes (the
so-called Extreme Mass Ratio Inspirals, or EMRIs). \citet{bc04}
present a detailed parameter estimation analysis for this class of
LISA events and show that they could be detected with good SNR out to
a distance $\sim 1$~Gpc. In addition, these authors found that the sky
localization errors for these events ($\sim 10^{-3}$ steradians at
$\sim 1$~Gpc) are comparable to the case of black hole binary merger
events. No electromagnetic counterpart is expected from EMRIs of dense
neutron stars or stellar-mass black holes. On the other hand, partial
(or total) disruptions of inspiraling low-density white dwarfs (WDs)
could produce such counterparts and thus help identify host galaxies
for such events. To the best of our knowledge, the possibility to find
the electromagnetic counterparts of this subclass of LISA EMRIs and
use them for precision cosmology has not been previously discussed in
the literature.

As a result of its relatively low density, a typical WD suffers
complete disruption around a non-spinning black hole of mass $\lsim
10^5 M_\odot$. This mass limit is increased to perhaps $\lsim 10^6
M_\odot$ if the black hole is spinning rapidly \citep{iv06,
rat05}. Even if the WD were not disrupted, some level of partial
shedding of its lower density outer layers \citep[e.g.][]{li02} may be
expected as the inspiral proceeds.  As the stream of debris from a
partially stripped WD shocks against itself or the inspiraling WD
\citep[e.g.,][]{ross08}, some level of electromagnetic emission during
such EMRIs is expected. A tidally-triggered detonation of the WD is
yet another possibility \citep{wilm04,dearb05}. All these arguments
point to the possibility that electromagnetic counterparts may exist
for a subset of all the EMRIs detected by LISA (the one corresponding
to WD inspirals). The fraction of all EMRIs which involve WDs is,
indeed, expected to be substantial \citep{ga04,ha06a,ha06b}.

If electromagnetic counterparts for a subset of all LISA EMRIs can be
detected, the rewards will be significant. Just like black hole binary
merger events, EMRIs with uniquely identified counterparts and host
galaxies could be used to draw a gravitational Hubble diagram, with
the significant advantages that, at redshifts as low as $z \lsim
0.2$--$0.3$, weak--lensing due to line-of-sight inhomogeneities would
be small or negligible (Fig.~\ref{fig:two}) and that the $D_L$--$z$
relation at these low redshifts is strongly sensitive to the dominant
dark energy content. Dark energy becomes dynamically significant,
affecting the expansion rate and geometry of the universe, and
modifying $D_L$, at $z\lsim 1$ \citep[see, e.g.][for a
review]{ht01}. Indeed, as emphasized by \citet{hh03}, once cosmic
microwave background anisotropies are measured by Planck, $D_L$ will
be known accurately at $z\sim 1000$, and a low--redshift ($z\sim 0$)
measurement will provide the best complement to constrain dark energy
parameters.

WD EMRIs detections with LISA will occur frequently, and they could
perhaps be singled out on the basis of the comparatively low mass of
the inspiraling compact object. It remains to be determined how the
LISA sky localization error evolves with pre-merger (or
pre-disruption) time for such events and what is the nature of their
electromagnetic counterparts. But altogether, the various advantages
that we have outlined point to the need for a detailed assessment of
the potential use of WD EMRIs for precision cosmology with LISA.

\section{Conclusion}

For centuries, astronomers have measured distances exclusively with
light.  Direct gravitational measurements, gravitational Hubble
diagrams and comparisons between the propagation of electromagnetic
and gravitational signals offer fundamentally new ways to probe
physics on cosmological scales. The novelty involved in joint,
time-constrained electromagnetic and gravitational measurements will
require that special efforts be made to reach out across the GW and
astronomy communities.

This work was supported by NASA grant NNX08AH35G. We thank
A. MacFadyen for the permission to use one of his figures.

% Bibliographic references with the natbib package:
% Parenthetical: \citep{Bai92} produces (Bailyn 1992).
% Textual: \citet{Bai95} produces Bailyn et al. (1995).
% An affix and part of a reference:
%   \citep[e.g.,][Ch. 2]{Bar76}
%   produces (e.g., Barnes et al. 1976, Ch. 2).

%\begin{deluxetable}{lrrrr} \tablecolumns{6}
%\tablewidth{0pt} \tablecaption{\label{tab:LISA}{\it LISA}
%Measurement Errors} \tablehead{\colhead{} & \colhead{$\delta\cal
%M/\cal M$} & \colhead{$\delta \mu/\mu$} & \colhead{$\delta \dL/\dL$}
%& \colhead{$\delta\Omega$}} \startdata
%   best     & $0.8\times10^{-5}$    &  $2\times10^{-5}$  &   $2\times10^{-3}$  
% &    $0.01 \deg^2$ \\
%   typical  & $2\times10^{-5}$      &  $9\times10^{-5}$  &   $4\times10^{-3}$  
% &    $0.3 \deg^2$  \\
%   worst    & $0.8\times10^{-3}$    &   0.1              &   $2\times10^{-2}$  
% &    $3 \deg^2$
%\enddata
%\tablecomments{Assumed SMBH binary parameters: $m_1=m_2=10^6 {\rm
%M_\odot}$ and $z=1$.}
%\end{deluxetable}

%\begin{figure}
%\centering \leavevmode
%\epsfxsize=\columnwidth \epsfbox{dbarret.ps}
%\caption{Measured quality factor of the lower (dots) and upper (squares) HFQPOs of
% 4U 1636-536. The solid  and dashed lines show the estimates done with the toy model
%proposed in \cite{bar06}, from which this figure is taken. }
%\label{fig:q}
%\end{figure}

% \bibitem[Names(Year)]{label} or \bibitem[Names(Year)Long names]{label}.
% (\harvarditem{Name}{Year}{label} is also supported.)

%\bibliographystyle{harvard}
%\bibliography{biblio}

%\begin{thebibliography}{}
%\bibitem[]{}
%\end{thebibliography}

\end{document}